\documentstyle[11pt]{article}
\input epsf

\newlength{\awidth}
\newlength{\aheight}
\newlength{\uswidth}
\newlength{\usheight}
\def\preprint#1{\gdef\@preprint{#1}}
\def\bce{\begin{center}}
\def\ece{\end{center}}
\def\be{\begin{equation}}
\def\ee{\end{equation}}
\def\bea{\begin{eqnarray}}
\def\eea{\end{eqnarray}}

\newcounter{fignr}
\newenvironment{fig}[1]{\refstepcounter{fignr}\label{#1}\begin{center}}{
    \end{center}}
\newcommand{\figcap}[2]{\parbox{#1}{{\footnotesize  Fig. \thefignr. #2}}}

\begin{document}
\baselineskip=.285in

\catcode`\@=11
\def\maketitle{\par
 \begingroup
 \def\thefootnote{\fnsymbol{footnote}}
 \def\@makefnmark{\hbox
 to 0pt{$^{\@thefnmark}$\hss}}
 \if@twocolumn
 \twocolumn[\@maketitle]
 \else \newpage
 \global\@topnum\z@ \@maketitle \fi\thispagestyle{empty}\@thanks
 \endgroup
 \setcounter{footnote}{0}
 \let\maketitle\relax
 \let\@maketitle\relax
 \gdef\@thanks{}\gdef\@author{}\gdef\@title{}\let\thanks\relax}
\def\@maketitle{\newpage
 \null
 \hbox to\textwidth{\hfil\hbox{\begin{tabular}{r}\@preprint\end{tabular}}}
 \vskip 2em \begin{center}
 {\Large\bf \@title \par} \vskip 1.5em {\normalsize \lineskip .5em
\begin{tabular}[t]{c}\@author
 \end{tabular}\par}
 \end{center}
 \par
 \vskip 1.0em}
\def\preprint#1{\gdef\@preprint{#1}}
\def\abstract{\if@twocolumn
\section*{Abstract}
\else \normalsize
\begin{center}
{\large\bf Abstract\vspace{-.5em}\vspace{0pt}}
\end{center}
\quotation
\fi}
\def\endabstract{\if@twocolumn\else\endquotation\fi}
\catcode`\@=12

\preprint{}
\title{\Large\bf Non-Newtonian Dynamic Gravitational Field from The Longitudinally 
Asymmetric Rotating Objects
\protect\\[1mm]\  }
\author{\normalsize Eue Jin Jeong\\[1mm]
{\normalsize\it Department of Physics, The University of Texas at Austin, 
Austin, TX 78712}}

\maketitle

\def\gatij{\gamma^{ij}}
\def\gabij{\gamma_{ij}}
\def\ophi{\phi^{a}}
\def\hphi{\overline{\phi}^{a}}
\def\dint{\int\!\!\!\!\!\int}
\begin{center}
{\large\bf Abstract}\\[3mm]
\end{center}
\indent\indent
\baselineskip=.285in

The dynamic shift of the center of mass for a rotating hemisphere 
prompts us the question of what might be its physical consequences.  
Despite the fact that accelerating object is known to create gravitational 
field, there is no known external dynamic gravitational force from a 
rotating sphere where the individual mass components are in constant 
acceleration.  However, Thirring's `induced centrifugal force' and 
the component of the force along the longitudinal axis inside a rotating 
spherical shell indicate that they are non-radiative dynamic forces 
which depend on $\omega^2$.  In this report, Thirring's force is derived by 
considering the component-wise acceleration of the rotating hemisphere 
in the weak field approximation.  This new analytic solution provides 
the gravitational explanation of the jet phenomena observed from the 
fast rotating cosmological bodies, which demands a major revision in 
our understanding of the universe since it suggests there exists a 
strong, long ranged, non-Newtonian dynamic gravitational force in 
our universe. This also raises an interesting question of how the 
strength of the dipole moment can be maximized for a given mass by 
configuring the specific geometrical shape of the rotating source.
\noindent
\newpage
\baselineskip=15pt
\pagenumbering{arabic}
\thispagestyle{plain}
\setcounter{section}{1}
\indent\indent
The problem of the non-Newtonian gravitational force experienced 
by a test particle inside a rotating spherical shell has been 
considered by Thirring[1] in 1918.  In his calculation within 
the weak field approximation, Thirring used the constant mass 
density r and the four velocity 
\\
\bea
u^1 &=& \frac{\omega R sin{\theta} sin{\phi}}{\sqrt{1 - \omega^2 R^2 
sin^{2}{\theta}}} \nonumber \\[5pt]
u^2 &=& \frac{\omega R sin{\theta} cos{\phi}}{\sqrt{1 - \omega^2 R^2 
sin^2{\theta}}} \nonumber \\[5pt]
u^3 &=& 0 \nonumber \\[5pt]
u^0 &=& \frac{1}{\sqrt{1 - \omega^2R^2 
sin^2{\theta}}} 
\eea
\\
and the length contraction 
\\
\be
d^3x' =d^3x''\sqrt {1 - \omega^2R^2 sin^2\theta}
\ee
\\
for the rotating spherical mass shell to perform the integration in 
the rest frame of the source to evaluate $\Phi_\mu^\nu$,
\\
\be
\Phi_\mu^\nu=4\int \frac{\rho u'_\mu u'^\nu d^3x'}
{ | \mathbf{r} - \mathbf{r}' | }
\ee
\\
from which $h_{\mu\nu}$ can be calculated as
\\
\be
h_{\mu\nu} = 
\Phi_{\mu\nu} - \frac{1}{2}\Phi
\ee
\\
In fact, one could have calculated the $\Phi_\mu^\nu$ in the rest frame of 
the observer by using the relativistic mass density in the same 
range of the radial integral and the resulting effects would 
have been the same since the integrand for $\Phi_\mu^\nu$ is the same for 
both cases.  By such a method of calculation[1], Thirring has 
effectively circumvented the problem of the questionable rigidity 
of the spherical mass shell.  On the other hand, physically, it 
is equivalent of taking the relativistic total mass-energy 
density $\gamma$($\omega,\theta$)$\rho$ for the dynamic mass 
components of the shell 
and then perform the integration in the observer's rest frame 
without concerning about the rigidity of the source.  

For a test particle located close to the center of mass of the 
rotating spherical mass shell of radius R with the angular 
frequency $\omega$, the Cartesian components of the acceleration 
(force/mass) have been shown to be given by, using the above 
method [1] [2] [3],
\\
\bea
\ddot{x}&=&\frac{M}{3R}(\frac{4}{5}\omega^2 x - 8\omega\nu_y) 
 \nonumber \\[5pt]
\ddot{y}&=&\frac{M}{3R}(\frac{4}{5}\omega^2 y + 8\omega\nu_x) 
 \nonumber \\[5pt]
\ddot{z}&=&-\frac{8M}{15R}\omega^2 z 
\eea
\\
In regard to this problem, Bass and Pirani[4] have reported earlier 
that the `induced centrifugal force' in the Thirring's geodesic 
equation[1][2][3] near the center of the rotating spherical mass 
shell arises as a consequence of the latitude dependent 
velocity distribution.  From this observation, Cohen and 
Sarill[5] suggested that the `induced centrifugal force' effect 
is due to the quadrupole moment.  To investigate this problem 
analytically, using the longitudinal symmetry of the problem, 
one may write
\\
\be
\Phi_\mu^\nu = 
4\int\frac{ \rho u'_\mu u'^\nu d^3 x' }{ | \mathbf{r} - \mathbf{r'} | }=
8 \pi \int_0^{\pi/2}\!\!\int \frac{\rho u'_\mu u'^\nu r'^2 sin \theta dr' 
d\theta'}{| \mathbf{r} - \mathbf{r'} |}+8\pi \int_{\pi/2}^{\pi}\!\!\int \frac{\rho u'_\mu u'^\nu 
r'^2 sin\theta dr' d\theta'}{|\mathbf{r} - \mathbf{r'}|}
\ee
\\
For $\omega R <<1$, the $T^{00}$ component of the stress energy tensor becomes 
the dominant term in Eq. (6). By employing the Eq.s (1) and (2), $\Phi_{00}$ for 
the upper half of the sphere may be written by
\\
\be
\Phi_{0(half)}^0 =
8\pi \int_0^{\pi/2}\!\!\int\frac{ \rho 
(\sqrt{1-\omega^2R^2sin^2\theta''})^{-1} sin \theta'' r''^2 dr''d\theta''}
{|\mathbf{r} - \mathbf{r}''|}
\ee
For the field outside the rotating hemispherical shell, the major 
contribution to the field from the half of the sphere in Eq. (7) 
is given by, up to the order $(1/r^2)$, 
\\
\be
\phi= 
-\frac{M}{r}-\frac{d_z}{r^2} cos\theta + O(\frac{1}{r^3}) 
\ee \\
where
\\
\be
d_z= 
M\delta r_c=MR \left( \frac{1}{2} -
\frac {\frac {1 - \sqrt{1 - \alpha}}{\alpha}}{\sqrt{\frac {1}{\alpha}}
\sinh^{-1}{\sqrt{\frac{\alpha}{1 - \alpha}}}} \right)
\ee
\\
\be
\alpha= 
\frac{\omega^2 R^2}{c^2}
\ee
\\
for the hemispherical shell (flat side down) of radius R, total 
mass M, rotating with the angular frequency $\omega$, where the origin 
of the coordinate system is located at the rest state $(\omega=0)$ center 
of mass of the hemisphere and $\delta r_c$ is defined as the anomalous center 
of mass shift.  The Eq.s (8) and (9) show that the strength of the 
dipole field depends on the magnitude of the anomalous shift of the 
center of mass which again depends on the geometrical configuration 
of the rotor.  When the two dipole fields are superposed inside a 
spherical shell, it turns out that the resulting non-Newtonian $\omega^2$
dependent gravitational force inside the spherical shell is formally 
identical to that of Thirring's `induced centrifugal force'[1][2][3].  
The major inverse $r^3$ forces from the two opposite dipole fields 
are canceled inside the sphere and there remain forces that behave 
like a harmonic oscillator in the z direction and that of a cascade 
in the radial direction.  These remaining $\omega^2$ dependent forces in 
the x, y, z directions close to the center of the sphere are given by 
\\
\bea
\ddot{x}&=&\frac{2M}{R}\omega^2 x 
\nonumber \\[5pt]
\ddot{y}&=&\frac{2M}{R}\omega^2 y 
\nonumber \\[5pt]
\ddot{z}&=&-\frac{4M}{R}\omega^2 z 
\eea
\\
Apart from the apparent formal resemblances, there are couple of 
discrepancies between this result and that of Thirring's.  The 
first conspicuous one is the difference in the constant factor of 2/15 
between the two expressions.  Also the information on the velocity 
dependent force is lost which is caused by the fact that the other 
components of the stress-energy tensor have been ignored except $T_{00}$ 
for the field outside of the source.  The discrepancy in the constant 
factor would have been expected since the position r=R/2 is not far 
outside of the boundary of the source, while the $1/|r-r'|$ 
expansion 
for the dipole moment was made with the assumption $r>r'=R/2$. 
Therefore, the multipole potential for a rotating spherical shell 
may be written
\\
\be
\phi= 
-\frac{M}{r}+\frac{d_z/2}{|-(R/2)\hat{z}-\textbf{r}|^2}cos\theta'-
\frac{d_z/2}{|(R/2)\hat{z}-\mathbf{r}|^2}cos\theta''+O(\frac{1}{r^3})
\ee
\\
where the angles $\theta'$ and $\theta''$ are given by
\\ 
\bea
\theta'&=& tan^{-1}\left( \frac{r sin\theta}{r cos\theta + R/2}\right)
\nonumber \\[5pt]
\theta''&=&tan^{-1}\left(\frac{r sin\theta}{r cos\theta - R/2}\right)
\eea
\\
respectively and $d_z$ is given by the Eq. (9).  The Eq. (12) is plotted 
in the three dimensional diagram of the potential in Fig. 1, which 
shows the quadrupole feature inside the rotating spherical mass shell 
as proposed earlier by Cohen and Sarill [5].

The problem is very similar to that of the electromagnetic vector 
potential from a circular ring of radius a with current I.  It is 
well known that the azimuthal component (the only non-zero term 
due to the symmetry) of the vector potential for both inside and 
outside of the radius a of the ring is approximately given by 
\\
\be
A_\phi(r,\theta)=\frac{\pi I a^2 r sin \theta}{c(a^2 + r^2)^{3/2}}
\left( 1 + \frac{15 a^2 r^2 sin^2 \theta}{8(a^2 + r^2)^2} + ...\right)
\ee
\\
For $r>>a$, the leading term of this potential depends on $1/r^2$ 
indicating the dipole effect.  It also gives the details of the 
potential inside the radius a without singularity.  Following this 
example, one may introduce a weight parameter $\eta$ into the gravitational 
dipole potential
\\
\be
\phi_{dipole}\propto\frac{-r}{(\eta^2 + r^2)^{3/2}} 
\ee
\\
so that the potential behaves without singularity for $r<r'$, where $\eta$ 
represents the parametrized radius of the physical object.  For a 
non-spherical body like a hemisphere, for example, one may assign 
the parameter tentatively a virtual physical dimension of a shell 
\\
\be
\eta=\sqrt{0.1} R
\ee
\\
which is about one third of the radius R of the sphere.  In this case, 
the corrected non-Newtonian force near the center of the sphere is 
reduced approximately by a factor 1/8 from the one in Eq. (11).  This 
is close to the value 2/15 which gives exactly Thirring's induced 
centrifugal force.  The above discussions suggest that the $w^2$ dependent 
forces in Thirring's result are mainly from the partially canceled 
dipole effect which arise due to the subtractive contribution to 
$F_z$ and the additive ones for $F_x$, $F_y$ from the two dipole moments 
respectively. 
In regard to this problem, Bass and Pirani[4] also have shown 
that the centrifugal force term arises as a consequence of the 
latitude dependent velocity distribution which generates an 
axially symmetric (non-spherical) mass distribution, which 
casts doubts on the centrifugal force interpretation of the 
Thirring's result since the rotating cylindrical object would 
not have such latitude dependent density distribution and 
there will be no corresponding centrifugal force for the cylindrical 
object, contrary to our expectation.  These difficulties remain 
even when the contribution from elastic stress is included, 
which led Bass and Pirani to conclude that there was an apparent 
conflict with Mach's principle.  Following this observation, 
Cohen and Sarill reported that the centrifugal term from Thirring's 
solution for a rotating spherical mass actually represents a 
quadrupole effect[5] by a deductive argument and suggested an 
alternative solution[6] (also previously by Pietronero[7]) for 
the centrifugal force in general relativity using the flat 
space metric in rotating coordinates. 
By employing the result in Eq. (15), the potential for a 
rotating spherical mass shell for both inside and out may 
be written, up to the dipole moment
\\
\be
\phi= 
V(r)+\frac{|-(R/2)\hat{z}-\textbf{r}|d_z/2}{\left(\eta^2+(-(R/2)
\hat{z}-\textbf{r})^2\right)^{3/2}}cos\theta'-\frac{|(R/2)\hat{z}-\textbf{r}|d_z/2}
{\left(\eta^2+((R/2)
\hat{z}-\textbf{r})^2\right)^{3/2}}cos\theta''+O(\frac{1}{r^3})
\ee
\\
where 
\\
\be
\begin{array}{lcr}
\begin{array}{lll}
 V(r) & = & -M/r \\
 & = & -M/R \\
\end{array} &
\mbox{\hspace{1cm}}&
\begin{array}{l}
 \textnormal{for $r>R$} \\
 \textnormal{for $r\leq R$} \\
\end{array}
\end{array}
\ee
\\
and the angles $\theta'$ and $\theta''$ are given by  Eq. (13).  
\\
\noindent\parbox{\textwidth}{\noindent\begin{fig}{fig1}
  \mbox{\setlength{\epsfxsize}{.80\textwidth} \epsfbox{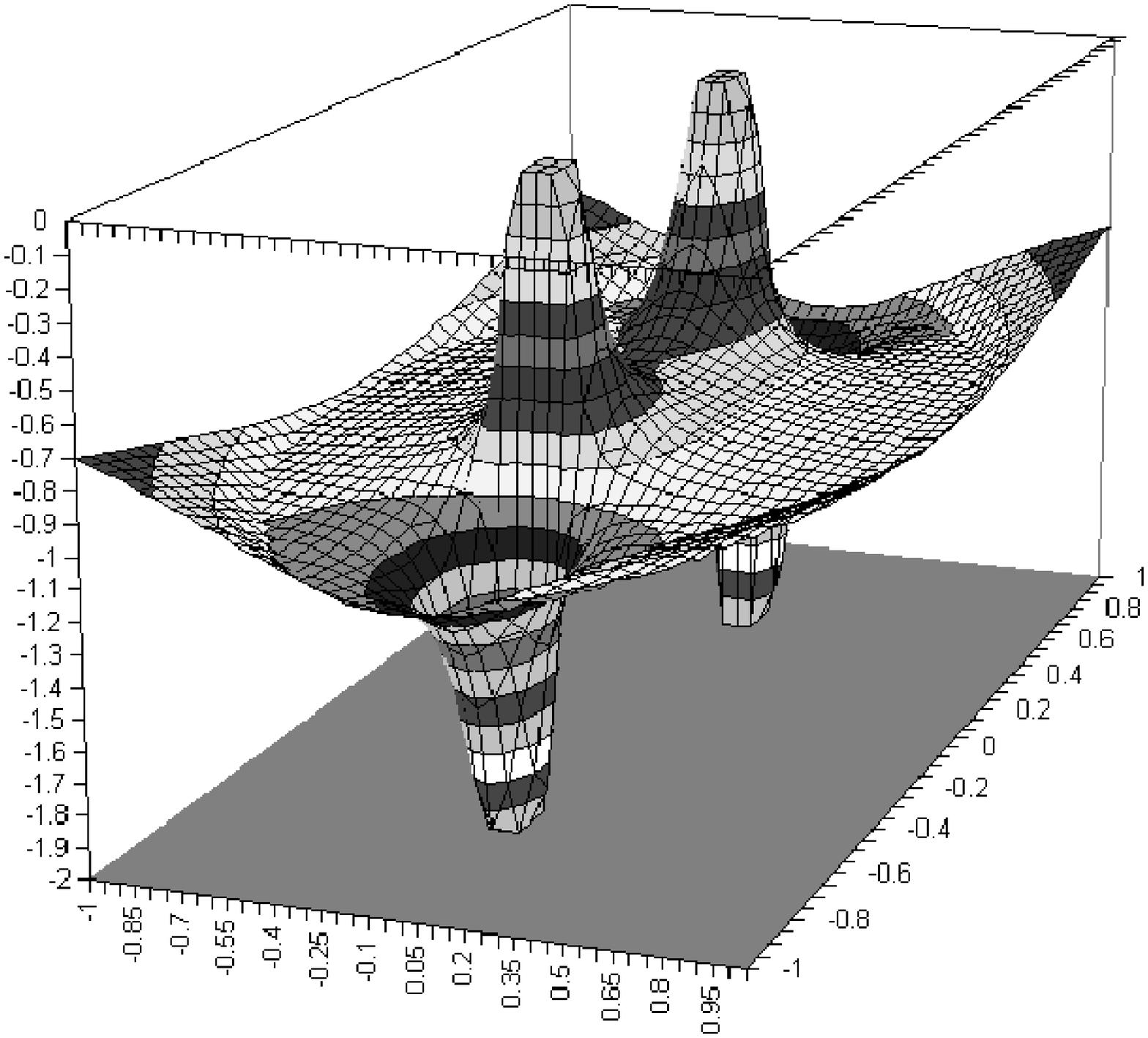} }\\
\ \\
  \figcap{.9\textwidth}{A close up view of the computer generated 3-D diagram 
for the multipole gravitational potential inside a spherical 
mass shell rotating along the z axis.  All the constants are 
set equal to 1 ($M=G=R=c=1$) and the value of the parameter $\eta$ 
is set equal to zero. The anomalous center of mass shift $\delta r_c$
for the potential in the diagram is 0.05R which corresponds 
to the case $\omega R=0.045 c$. The variable range of $\delta r_c$ 
is from 0 to 0.5R.}
\end{fig}
}\\
This potential is  plotted in Fig. 1 for a close up view of the inside field of the 
sphere for $\omega R=0.05c$.  The actual potential inside the rotating 
sphere must include the higher order terms which are similar to 
the ones in Eq. (14) to correctly represent the inner potential.  
In general, the behavior of the potential in the longitudinal axis  
suggests (Fig. 2) the possibility that the particles traveling into 
the attractive dipole potential well along the z axis will be repelled 
back to where they have come from depending on the rotational 
frequencies that support the height of the peaks. This repulsive 
potential peak determines the range of the linear orbital distances 
that the particles may travel back and forth from the poles to the 
far outsides along the z axis.
 
Since there is no compelling evidence 
that the plasma and magnetic field must be generated inside rotating 
ultra-compact bodies, where the electronic orbital states have been 
long before collapsed, one may suspect that the superposed dipole 
effect may have been the major driving force behind the jet 
phenomena in some of the fast rotating cosmological objects.
\\
\noindent\parbox{\textwidth}{\noindent\begin{fig}{fig2}
  \mbox{\setlength{\epsfxsize}{.78\textwidth} \epsfbox{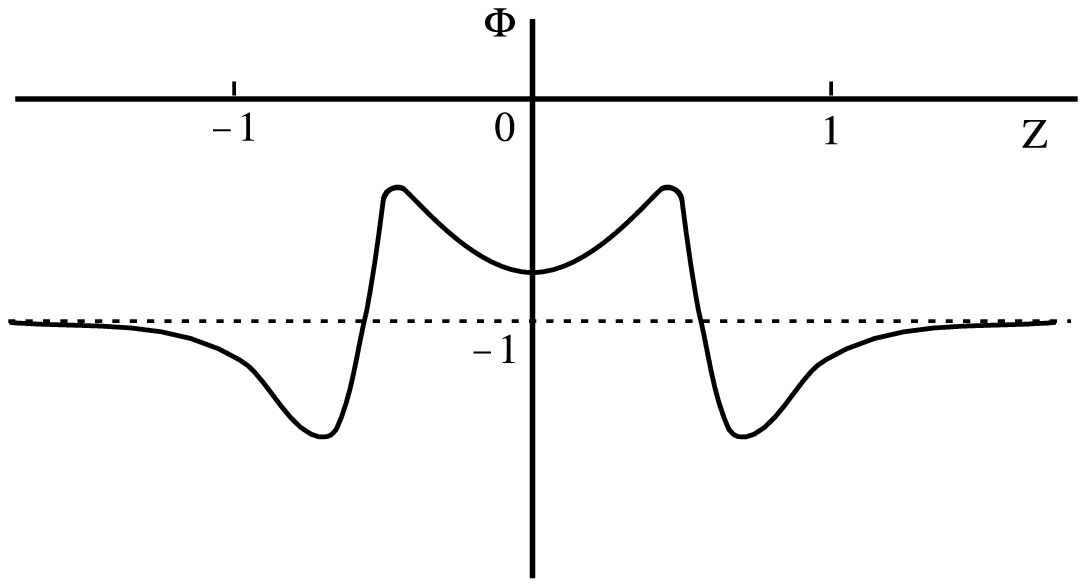} } 
 
\vspace{-198pt}
  \figcap{.9\textwidth}{An example of the superposed dipole field 
  (Eq. 17) along the longitudinal (z) axis 
  ($\eta^2=0.02$, M=G=c=R=1, $\omega R=0.05c$) }
\end{fig}
}\\   
Collisions and vibrations of particles coming in and going out following the linear 
trajectories may generate wide range of frequencies of radiations observed from 
quasars.
 This view point is supported by the fact that the dipole field 
is long ranged and the strongest next to that of the monopole.  
The long range potential dip around the equator in the diagram also 
indicates that there exists a tendency of the cluster formation 
around the equatorial plane of rotating celestial bodies. 
Since the dipole field depends on the geometrical shape of 
the rotor, it is interesting to find out what will be the most 
efficient geometrical configuration a rotor could take for given 
mass to produce the maximum dipole moment.  This question is 
interesting since the object under the influence of multipole 
gravitational potential will tend to take a shape which is 
different from the spherical symmetry.  The strength of the 
dipole moment defined by $M \delta r_c$ may be treated as an independent 
gravitational charge that determines the gravitational dipole 
field around the object the same way as one determines the 
electric or magnetic dipole moment.  Since this quantity depends 
critically on the longitudinal asymmetry of the rotor, one may 
start by assuming that the contour of the rotor assumes the 
form $z=a|x|^b$, where z is the rotation axis.  

To be able to 
determine the relative efficiency of the dipole rotor, a 
constraint must be imposed on the total rest mass of the rotor.  
Thus, the problem reduces into mathematically determining 
the geometrical contour that gives the maximum shift of the 
center of mass for a given mass of the rotor.
The total rest mass of the solid object, the contour of which 
is expressed by the equation $z=a|x|^b$ (for $-x_s<x<x_s$, $a>0$, $b>0$), 
is written in the rotor's rest frame by
\\
\be
M_0=\rho V=\rho \int \pi abx^{b+1}dx=\rho \frac{\pi ab}{b+2}x_s^{b+2}
\ee
\\
where $x_s$ is the radius of the widest area of the object in the 
plane of rotation and r is the density of the material.  The z 
component of the dipole moment of the object rotating with the 
angular frequency $\omega$ along the z axis is given by, in the observer's 
frame, following the method of Thirring's for $r>r'$( note that x, y 
components are zero due to the symmetry),
\\
\be
Mz=\int \rho\pi a b x^{b+1}ax^b \frac{dx}{\sqrt{1-\frac{\omega^2x^2}{c^2}}}
\ee
\\
Since we are interested in the case for $\omega x<<c$, the integral can 
be approximated to be 
\\
\be
Mz=\int \rho\pi a^2 bx^{2b+1}\left( 1+\frac{\omega^2x^2}{2c^2}\right) dx
\ee
\\
which gives the result
\\
\be
Mz\approx\rho\pi a^2b \left(\frac{1}{2b+2}x_s^{2b+2}+\frac
{\omega^2}{2c^2}\frac{1}{2b+4}x_s^{2b+4}\right)
\ee
\\
where the first term is the fictitious dipole moment of the rotor.  
This term can be eliminated by the coordinate translation.  
The second term is due to the latitude dependent velocity distribution 
which causes the coordinate independent, anomalous shift of the center 
of mass that may be written, using the Eq. (19), as
\\
\be
d_z=M\delta_z\approx\frac{\omega^2M_0^2}{4\pi\rho c^2}(1+\frac{2}{b})
\ee
\\
For instance, for a rotating cylindrical mass shell, the $\omega x$ in 
Eq. (20) would have to be replaced by $\omega x_0$ where $x_0$ is the radius 
of the cylinder.  Consequently, the $\gamma$ factor in Eq. (20) becomes 
a constant and the corresponding expression for Eq. (22) would 
not have the second term which represents the gravitational 
dipole moment.  This is in accordance with the observation 
reported by Bass and Pirani [4] that Thirring's `induced 
centrifugal force' is actually due to the latitude dependent 
velocity distribution which does not exist in a rotating cylindrical mass.
The above result indicates that the smaller the value of b, 
the greater the efficiency of the dipole effect.  However, 
the value of b affects the other constraint since the total 
rest mass of the rotor depends both on the values of a and b 
respectively by the relation (19). 
 
\vspace{10pt}
\noindent\parbox{\textwidth}{\noindent\begin{fig}{fig3}
  \mbox{\setlength{\epsfxsize}{.65\textwidth} \epsfbox{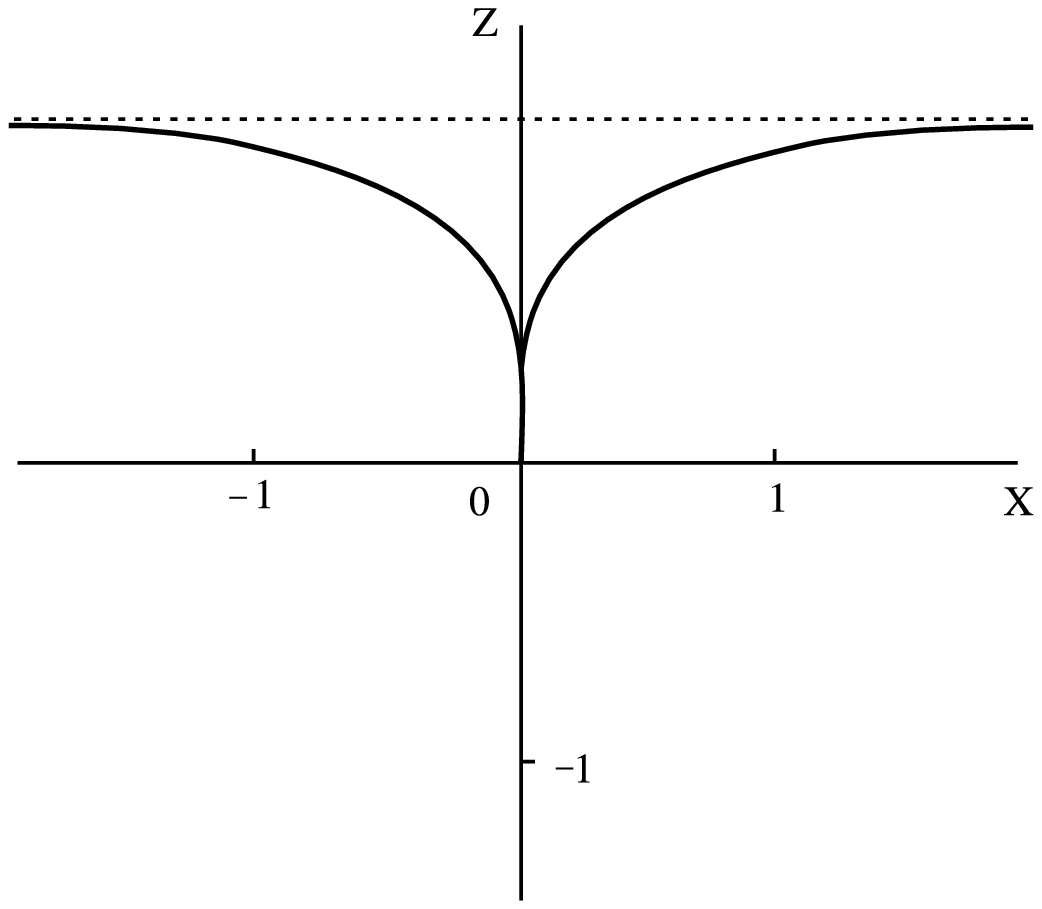} } \\
\ \\
\figcap{.9\textwidth}{The shape of an efficient dipole rotor 
  which has the contour equation \break $z=|x|^{1/5}$}
\end{fig}
}\\ 

By assuming that b is 
much smaller than 1, the expression (19) can be rewritten
\\
\be
M_0\approx\rho\frac{\pi ab}{2}x_s^2
\ee
\\
The smaller b requires correspondingly the larger a to 
maintain the constant rest mass.  The Eq. (23) also suggests 
that the lower the density of the material comprising the rotor, 
the greater the effect of the dipole moment when the constraint 
is on the mass of the rotor, in which case a larger $x_s$ would be 
required.  In general, the form of the curve for small b $(b<1)$ 
suggests that the outer shape of the rotor must be like a cusped 
funnel, the contour of which is plotted in Fig. 3,  in the first 
order approximation to have the efficient dipole moment effect.  
For extremely small b, the side view of the contour becomes the 
shape of a letter T, in which case the efficiency may approach 
the maximum.  An interesting observation in regard to this 
problem is that a fast rotating galaxies tend to have the shape of 
two superposed opposite funnels (Fig. 4).  
\\
\noindent\parbox{\textwidth}{\noindent\begin{fig}{fig4}
  \mbox{\setlength{\epsfxsize}{.80\textwidth} \epsfbox{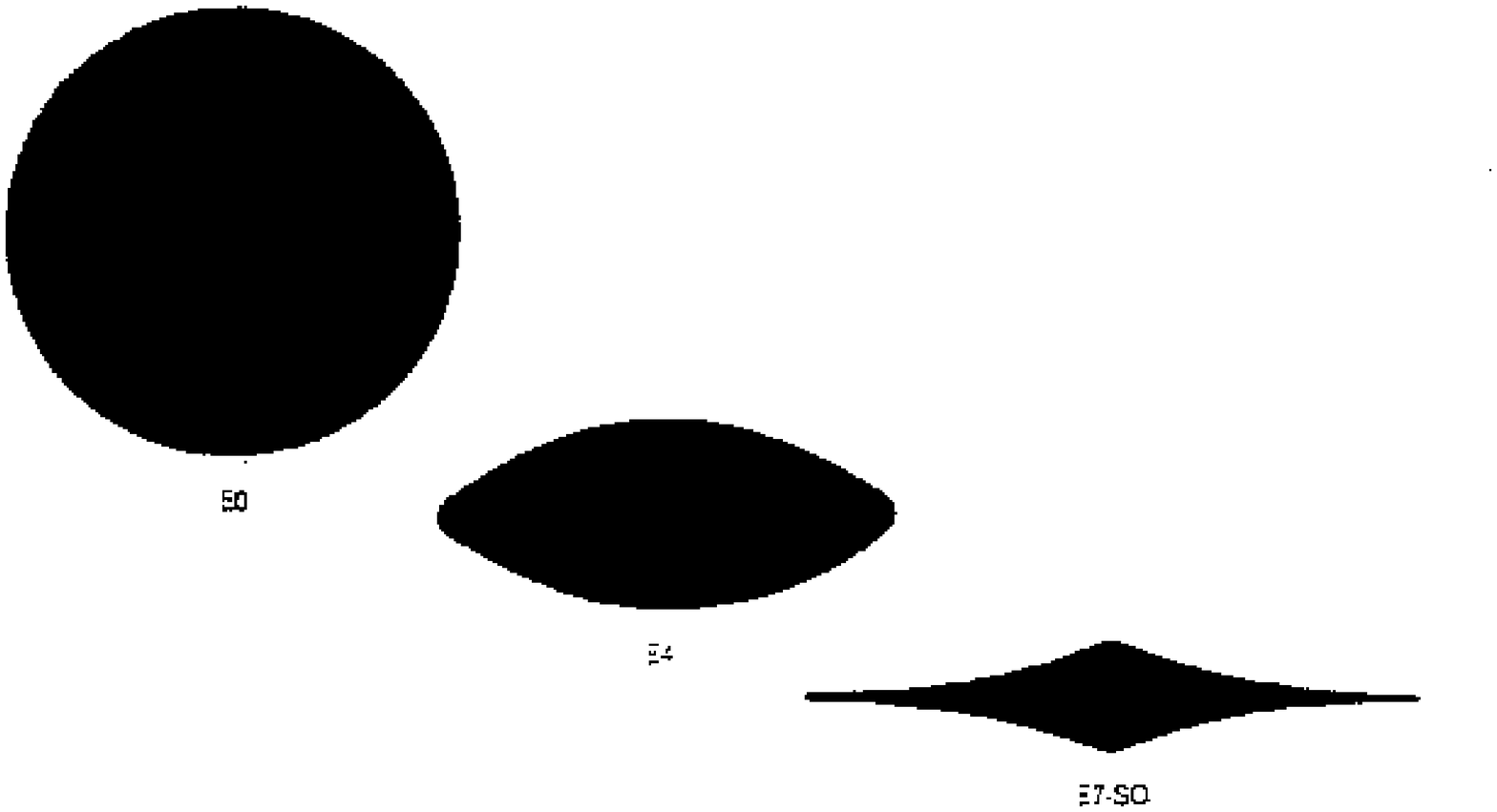} }\\
\ \\
  \figcap{.9\textwidth}{The shape of the galaxies in the order of 
increasing rotational frequencies.}
\end{fig}
}\\ 
It seems that 
the cosmological objects find the most efficient contour by 
itself, depending on the rotational frequencies of the whole 
body.  In the first order, one observes the celestial bodies 
take the form of a spherical shape due to the isotropy of the 
monopole field and in the second order, when the fast rotational 
motion is involved, the cosmological bodies transform into a 
longitudinal axially symmetric shape like that of a superposed 
funnel or a disc with long jets.  As can be seen from  Fig. 2, 
mass tends to be accumulated near the poles which would transform 
the spherical mass into a cusped shape which would again help 
increase the efficiency of the dipole moment.  This process is 
self perpetuating and suggests a strong possibility that the 
jets in the black hole accretion discs are an extended form of the 
general configuration of the fast rotating cosmological bodies 
resulting from the dynamic gravitational dipole field.

\def\hebibliography#1{\begin{center}\subsection*{References}
\end{center}\list
  {[\arabic{enumi}]}{\settowidth\labelwidth{[#1]}
\leftmargin\labelwidth	  \advance\leftmargin\labelsep
    \usecounter{enumi}}
    \def\newblock{\hskip .11em plus .33em minus .07em}
    \sloppy\clubpenalty4000\widowpenalty4000
    \sfcode`\.=1000\relax}

\let\endhebibliography=\endlist

\begin{hebibliography}{100}

\bibitem{B1} H. Thirring, Z. Phys., {\bf 19}, 33 (1918); and {\bf 22}, 29 (1921)
\bibitem{B2} J.  Weber, General Relativity and Gravitational Waves 
(Interscience Publisher, Inc., New York, 1961), p. 160
\bibitem{B3} L. D. Landau and E.  Lifshitz, The Classical Theory of 
Fields (Addison-Wesley Publishing Company, Inc., Reading, Massachusetts, 1959)
\bibitem{B4} Bass, L., and Pirani, F.  A.  E., Phil.  Mag., {\bf 46}, 850 (1955)
\bibitem{B5} J.  M.  Cohen and W.  J.  Sarill, Nature, {\bf 228}, 849 (1970)
\bibitem{B6} J. M.Cohen, W. J. Sarill  and C. V. Vishveshwara, 
Nature, {\bf 298}, 829 (1982)
\bibitem{B7} L. Pietronero, Ann. Phys., {\bf 79}, 250(1973)

\end{hebibliography}
\end{document}